\RequirePackage{lineno}

\documentclass[aps,prc,twocolumn,groupedaddress,amsmath,amssymb,floatfix,superscriptaddress,showpacs,altaffilletter,altaffilsymbol]{revtex4}
\usepackage{multirow}
\usepackage{graphicx}
\usepackage{times}
\usepackage{color}

\begin{document}

\bibliographystyle{h-physrev3}

\title{$^{7}$Be Solar Neutrino Measurement with KamLAND\\
}

%

\newcommand{\tohoku}{\affiliation
	{Research Center for Neutrino Science, Tohoku University, 
	Sendai, Miyagi 980-8578, Japan}}
\newcommand{\osaka}{\affiliation
	{Graduate School of Science, Osaka University, 
	Toyonaka, Osaka 560-0043, Japan}}
\newcommand{\alabama}{\affiliation
	{Department of Physics and Astronomy, University of Alabama, 
	Tuscaloosa, Alabama 35487, USA}}
\newcommand{\lbl}{\affiliation
	{Physics Department, University of California, Berkeley, and \\ Lawrence Berkeley National Laboratory, 
	Berkeley, California 94720, USA}}
\newcommand{\colostate}{\affiliation
	{Department of Physics, Colorado State University, 
	Fort Collins, Colorado 80523, USA}}
\newcommand{\hawaii}{\affiliation
	{Department of Physics and Astronomy, University of Hawaii at Manoa, 
	Honolulu, Hawaii 96822, USA}}
\newcommand{\tennessee}{\affiliation
	{Department of Physics and Astronomy, University of Tennessee, 
	Knoxville, Tennessee 37996, USA}}
\newcommand{\tunl}{\affiliation
	{Triangle Universities Nuclear Laboratory, 
	Durham, North Carolina 27708, USA and \\
	Departments of Physics at Duke University, North Carolina Central University,
	and the University of North Carolina at Chapel Hill}}
\newcommand{\ipmu}{\affiliation
	{Kavli Institute for the Physics and Mathematics of the Universe (WPI), University of Tokyo, 
	Kashiwa, Chiba 277-8583, Japan}}
\newcommand{\nikhef}{\affiliation
	{Nikhef and the University of Amsterdam, Science Park, 
	Amsterdam, the Netherlands}}
\newcommand{\washington}{\affiliation
	{Center for Experimental Nuclear Physics and Astrophysics, University of Washington, 
	Seattle, Washington 98195, USA}}
\newcommand{\caltech}{\affiliation
	{Kellogg Radiation Laboratory, California Institute of Technology,
	Pasadena, California 91125 USA}}
\newcommand{\drexel}{\affiliation
	{Department of Physics, Drexel University,
	Philadelphia, Pennsylvania 19104, USA}}
\newcommand{\kansas}{\affiliation
	{Department of Physics, Kansas State University,
	Manhattan, Kansas 66506, USA}}
\newcommand{\wisconsin}{\affiliation
	{Department of Physics, University of Wisconsin, 
	Madison, Wisconsin 53706, USA}}
\newcommand{\stanford}{\affiliation
	{Department of Physics, Stanford University,
	Stanford, California 94305, USA}}

\newcommand{\aticrrnow}{\altaffiliation
	{Present address (The same shall apply hereinafter.): \\ 
	~~Kamioka Observatory, Institute for Cosmic Ray Research, the University of Tokyo,
	 Hida, Gifu 506-1205, Japan}}
\newcommand{\atipmunow}{\altaffiliation
	{Kavli Institute for the Physics and Mathematics of the Universe (WPI), University of Tokyo, 
	 Kashiwa, Chiba 277-8583, Japan}}
\newcommand{\atosakanow}{\altaffiliation
	{Department of Physics, Osaka University, 
	Toyonaka, Osaka 560-0043, Japan}}
\newcommand{\atrcnpnow}{\altaffiliation
	{Graduate School of Engineering, University of Fukui, 
	Fukui 910-8507, Japan}}  
\newcommand{\atdavisnow}{\altaffiliation
	{Physics Department, University of California, Davis, 
	Davis, California 95616, USA}}
\newcommand{\atlivermorenow}{\altaffiliation
	{Lawrence Livermore National Laboratory,
	Livermore, California 94550, USA}}
\newcommand{\atfermilabnow}{\altaffiliation
	{Fermi National Accelerator Laboratory,
	Batavia, Illinois 60510, USA}}
\newcommand{\atmitnow}{\altaffiliation 
	{Department of Physics, Massachusetts Institute of Technology, Cambridge,
	Massachusetts 02139, USA}}
\newcommand{\atlosalamosnow}{\altaffiliation
	{Los Alamos National Laboratory,
	Los Alamos, New Mexico 87545, USA}}
\newcommand{\atjeffersonlabnow}{\altaffiliation
	{Thomas Jefferson National Accelerator Facility,
	Newport News, Virginia 23606, USA}}
\newcommand{\atbrookhavennow}{\altaffiliation
	{Brookhaven National Laboratory, 
	Yaphank, New York 11980, USA}}
\newcommand{\atyalenow}{\altaffiliation
	{Department of Physics, Yale University, 
	New Haven, Connecticut 06520, USA}}

\newcommand{\atmissourinow}{\altaffiliation
	{Chemical and Physical Sciences Department, Missouri Southern State University,
	 Joplin, Missouri 64801, USA}}
\newcommand{\atcarletonnow}{\altaffiliation
	{Department of Physics, Carleton University,
	Ottawa, Ontario K1S 5B6, Canada}}
\newcommand{\atlouisiananow}{\altaffiliation
	{Department of Physics and Astronomy, Louisiana State University,
	Baton Rouge, Louisiana 70803, USA}}
\newcommand{\atkeknow}{\altaffiliation
	{Institute of Particle and Nuclear Studies, High Energy Accelerator Research Organization,
	Tsukuba, Ibaraki 305-0801, Japan}}
\newcommand{\atarcadianow}{\altaffiliation
	{Department of Chemistry and Physics, Arcadia University,
	Glenside, Pennsylvania 19038, USA}} 
\newcommand{\atchinahongkongnow}{\altaffiliation
	{Department of Physics, the Chinese University of Hong Kong, 
	Shatin NT, Hong Kong SAR, China}}
\newcommand{\atlblnow}{\altaffiliation
	{Lawrence Berkeley National Laboratory, 
	Berkeley, California 94720, USA}}
\newcommand{\atwashingtonnow}{\altaffiliation
	{Department of Physics, University of Washington, 
	Seattle, Washington 98195, USA}}
\newcommand{\deceased}{\altaffiliation {Deceased.}}
\newcommand{\atsomewherenow}{\altaffiliation
	{\textcolor{red}{Unknown Current Affiliation}}}

\newcommand{\OscPhaseEff}{\Delta m^{2}\mathbb{L}/\mathbb{E})_{\rm{Eff}}}

%
%

\author{A.~Gando}\tohoku
\author{Y.~Gando}\tohoku
\author{H.~Hanakago}\tohoku
\author{H.~Ikeda}\tohoku
\author{K.~Inoue}\tohoku\ipmu
\author{K.~Ishidoshiro}\tohoku
\author{H.~Ishikawa}\tohoku
\author{Y.~Kishimoto}\aticrrnow\tohoku\ipmu
\author{M.~Koga}\tohoku\ipmu
\author{R.~Matsuda}\tohoku
\author{S.~Matsuda}\tohoku
\author{T.~Mitsui}\tohoku
\author{D.~Motoki}\tohoku
\author{K.~Nakajima}\atrcnpnow\tohoku
\author{K.~Nakamura}\tohoku\ipmu
\author{A.~Obata}\tohoku
\author{A.~Oki}\tohoku
\author{Y.~Oki}\tohoku
\author{M.~Otani}\atkeknow\tohoku
\author{I.~Shimizu}\tohoku
\author{J.~Shirai}\tohoku
\author{A.~Suzuki}\tohoku
\author{K.~Tamae}\tohoku
\author{K.~Ueshima}\tohoku
\author{H.~Watanabe}\tohoku
\author{B.D.~Xu}\tohoku
\author{S.~Yamada}\atkeknow\tohoku
\author{Y.~Yamauchi}\tohoku
\author{H.~Yoshida}\atosakanow\tohoku

\author{A.~Kozlov}\ipmu
\author{Y.~Takemoto}\ipmu

\author{S.~Yoshida}\osaka

\author{C.~Grant}\atdavisnow\alabama
\author{G.~Keefer}\atlivermorenow\alabama
\author{D.W.~McKee}\atmissourinow\alabama
\author{A.~Piepke}\alabama\ipmu

\author{T.I.~Banks}\lbl
\author{T.~Bloxham}\lbl
\author{S.J.~Freedman}\deceased\lbl
\author{B.K.~Fujikawa}\lbl\ipmu
\author{K.~Han}\lbl
\author{L.~Hsu}\atfermilabnow\lbl
\author{K.~Ichimura}\aticrrnow\lbl
\author{H.~Murayama}\lbl\ipmu
\author{T.~O'Donnell}\lbl
\author{H.M.~Steiner}\lbl
\author{L.A.~Winslow}\atmitnow\lbl

\author{D.~Dwyer}\atlblnow\caltech
\author{C.~Mauger}\atlosalamosnow\caltech
\author{R.D.~McKeown}\atjeffersonlabnow\caltech
\author{C.~Zhang}\atbrookhavennow\caltech

\author{B.E.~Berger}\colostate\ipmu

\author{C.E.~Lane}\drexel
\author{J.~Maricic}\drexel\hawaii
\author{T.~Miletic}\atarcadianow\drexel

\author{J.G.~Learned}\hawaii
\author{M.~Sakai}\hawaii

\author{G.A.~Horton-Smith}\kansas
\author{A.~Tang}\kansas

\author{K.E.~Downum}\stanford
\author{K.~Tolich}\atwashingtonnow\stanford\washington

\author{Y.~Efremenko}\tennessee\ipmu
\author{Y.~Kamyshkov}\tennessee
\author{O.~Perevozchikov}\atlouisiananow\tennessee

\author{H.J.~Karwowski}\tunl
\author{D.M.~Markoff}\tunl
\author{W.~Tornow}\tunl\ipmu

\author{J.A.~Detwiler}\washington
\author{S.~Enomoto}\washington\ipmu

\author{K.~Heeger}\atyalenow\wisconsin

\author{M.P.~Decowski}\nikhef\ipmu

\collaboration{The KamLAND Collaboration}\noaffiliation

%


\begin{abstract}
	We report a measurement of the neutrino-electron elastic scattering rate of 862\,keV $^{7}$Be solar neutrinos based on 
	a 165.4\,kton-days exposure of KamLAND. The observed rate is \mbox{$582 \pm 94\,($kton-days$)^{-1}$}, 
	which corresponds to a 862\,keV $^{7}$Be solar neutrino flux of $(3.26 \pm 0.52) \times 10^{9}\,{\rm cm}^{-2}{\rm s}^{-1}$, 
	assuming a pure electron-flavor flux. 
	Comparing this flux with the standard solar model prediction and further assuming three-flavor mixing, 
	a $\nu_{e}$ survival probability of $0.66 \pm 0.15$ is determined from the KamLAND data. 
	Utilizing a global three-flavor oscillation analysis, we obtain a total $^{7}$Be solar neutrino flux of 
	$(5.82 \pm 1.02) \times 10^{9}\,{\rm cm}^{-2}{\rm s}^{-1}$, which is consistent with the standard solar model predictions.
\end{abstract}


\maketitle

%

\section{Introduction}
\label{section:Introduction}

	\vspace{-0.2cm}
	
	During the past 40 years, solar neutrino flux measurements have been obtained
	through the use of a wide range 
	of detection techniques and neutrino interactions~\cite{Cleveland1998, Hampel1999, Altmann2005, 
	Abdurashitov2009, Hosaka2006, Aharmim2010, Aharmim2008, Arpesella2008b}.
	This research tested the validity of the solar fusion model and provided compelling evidence
	for matter induced neutrino flavor conversion within the solar interior,
	called the MSW effect~\cite{Wolfenstein1979, Mikheyev1985}.  
	The observation of anti-neutrino flavor oscillations by the 
	KamLAND experiment~\cite{Gando2013}, having mixing parameters consistent with
	those needed to explain the solar data, showed neutrino mixing
	to be responsible for what had been known as the {\it Solar Neutrino Problem}.
	
	The most recent solar neutrino experiments are liquid scintillator 
	based~\cite{Bellini2013,Abe2011}. The low energy threshold possible with
	such detectors allows for real-time detection of the low energy $^7$Be solar neutrinos.
	Neutrinos from $^{7}$Be electron-capture inside the sun are mono-energetic (862\,keV) 
	and dominate the solar neutrino spectrum above the energies of $pp$ 
	neutrinos ($<$ 420\,keV). The flux of $^{7}$Be solar neutrinos
	has been previously measured by the Borexino experiment~\cite{Bellini2013}.
	In this article, we report a measurement of  the $^7$Be solar neutrino flux with KamLAND, 
	thereby providing the first independent cross-check of this important quantity.
	
	As in Borexino, solar neutrinos are detected via neutrino-electron elastic scattering, 
	$\nu + e \rightarrow \nu + e$ (ES), 
	which has a well understood cross section. In the standard three-flavor mixing scheme, 
	electron neutrinos (${\nu}_{e}$) produced inside the sun can transform 
	into muon or tau neutrinos (${\nu}_{\mu}$ or ${\nu}_{\tau}$) during
	flight at a rate determined by the neutrino oscillation parameters and the electron density of the solar interior; 
	the total active neutrino flux (${\nu}_{e}$ + ${\nu}_{\mu}$ + ${\nu}_{\tau}$) is conserved. 
	At the energy of $^{7}$Be solar neutrinos, the ES cross section of ${\nu}_{e}$ 
	is about five times larger than that of ${\nu}_{\mu}$ or ${\nu}_{\tau}$. 
	
	The expected $^7$Be neutrino flux on the earth's surface, given by the GS98 solar model~\cite{Grevesse1998}, 
	is $5.00 \times 10^{9}\,{\rm cm}^{-2}{\rm s}^{-1}$~\cite{Serenelli2011}.
	The necessity of a large detector is evident considering the small ES cross section, which gives, for the KamLAND scintillator, 
	an expected interaction rate of $500\,($kton-days$)^{-1}$ including neutrino flavor conversion. 
	The ES induced recoil electrons have to be measured 
	without the benefit of a convenient event tag.  Here lies the main difficulty of these experiments: 
	the low-energy backgrounds of a kiloton-size detector have to be sufficiently suppressed to allow
	the observation of a signal composed of only few events, a non-trivial task. In the current work this is
	achieved by comparing detector background models with and without
	a solar recoil signal to the data. The presence of such a signal is then
	inferred from the data by means of a chi-square statistical analysis.

%

\section{Detector and Calibration}
\label{section:Detector}

	\vspace{-0.2cm}

\begin{figure}[t]
		\begin{center}
			\includegraphics[angle=270,width=1.0\columnwidth]{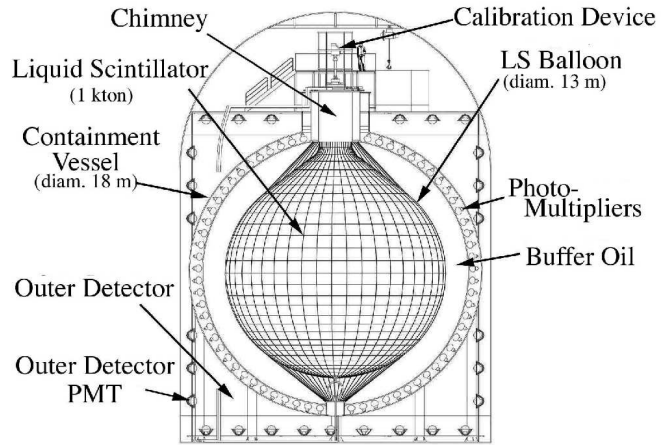}
			\vspace{-0.8cm}
		\end{center}
		\caption[]{Schematic diagram of the KamLAND detector.}
		\label{figure:detector}
	\end{figure}

	The KamLAND detector (Figure\,\ref{figure:detector}) consists of 1\,kton of liquid scintillator  (LS) contained 
	in a thin plastic-film balloon of 13 m diameter. The scintillation light is viewed by an array of 1879 
	photomultiplier tubes (PMTs) mounted on the inner surface of an 18 m diameter stainless steel sphere (SSS). 
	The space between the SSS and the balloon is filled with purified mineral oil which shields the LS from external radiation.
	The SSS and its content, denoted the inner detector (ID), is contained within a cylindrical, 3.2\,kton water-Cherenkov 
	outer detector (OD).
	All detector materials and components were selected to have low 
	radioactivity content to maintain the option of a low background phase. 
	
	The KamLAND detector started collecting data for the reactor anti-neutrino phase in March 2002. 
	Due to the delayed coincidence structure (a prompt positron followed by a delayed neutron capture)
	with which anti-neutrinos can be tagged the background was low enough for their detection.
	However the analysis of low energy singles data (composed of events not benefitting from a delayed coincidence)
	showed that the liquid scintillator contained $883 \pm 20\,\mu {\rm Bq/kg}$ of $^{85}$Kr and
	$58.4 \pm 1.1\,\mu {\rm Bq/kg}$ of $^{210}$Pb, the latter inferred from the
	decay rates of its unstable daughters $^{210}$Bi and $^{210}$Po.
	The resulting total decay rate of $8.1 \times 10^{7}\,($kton-days$)^{-1}$ 
	made the detection of about $500\,($kton-days$)^{-1}$ $^7$Be solar neutrino
	induced recoil electrons impossible.

	In order to enable a low energy physics program with KamLAND, the collaboration
	developed methods for the efficient removal  of Kr and Pb from the 
	liquid scintillator~\cite{Keefer2015}. Based on extensive small-scale laboratory studies, 
	large scintillator distillation and nitrogen purge systems were constructed underground.
	Two purification campaigns were performed in 2007 and 2008-2009.  During the purification
	campaigns the old LS was drained from the detector and simultaneously filled with recycled, purified LS 
	such that the LS mass supported by the balloon remained constant.  To maximize the efficiency of
	purification, the temperature and density of the purified LS was carefully controlled in order to 
	maintain a boundary between the old and purified LS. By the end of both campaigns more 
	than five detector volume exchanges were performed, resulting
	 in a substantial reduction of the background-creating impurities.  
	The overall reduction factors for rates of $^{85}$Kr, $^{210}$Bi, and $^{210}$Po,
	were about $6 \times 10^{-6}$, $8 \times 10^{-4}$, and $5 \times 10^{-2}$, respectively. 
	This dramatic reduction allowed the primary trigger threshold 
	to be lowered from 180 PMT hits to 70 PMT hits (the latter value corresponds 
	to a threshold of $\sim$0.4\,MeV), and thus extended KamLAND's
	scientific reach into the detection of low energy solar neutrinos.
	To allow more detailed study of low energy backgrounds, the threshold is lowered once per second to $\sim$0.2\,MeV for a duration of 1\,ms.
	The data presented were collected in 616 days between April 7, 2009 and June 21, 2011. 
	
\begin{figure}[t]
		\begin{center}
			\includegraphics[angle=0,width=1.0\columnwidth]{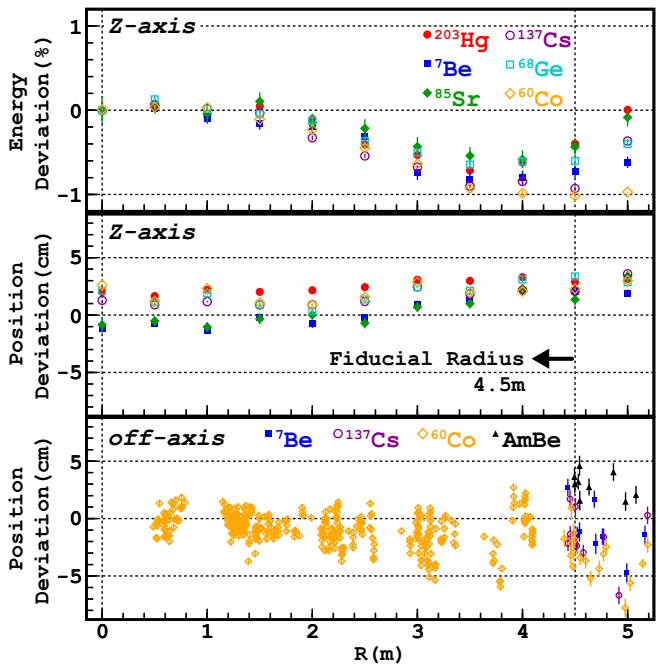}
			\vspace{-0.8cm}
		\end{center}
		\caption[]{"Color Online" Deviation of the reconstructed energy with respect to z = 0~m and deviation of the reconstructed position 
		with respect to the true position for gamma calibration sources inside the detector.}
		\label{figure:z-axis-calibration}
	\end{figure}

	The event position and energy are reconstructed based on the time and charge of 
	photon-hits recorded by the PMTs. The KamLAND coordinate system utilizes the horizontal equatorial plane as its xy-plane; 
	the z-axis points up.  The reconstruction is calibrated using gamma sources deployed periodically in the detector --- namely 
	$^{7}$Be (0.478 MeV), $^{60}$Co (2.506 MeV), $^{68}$Ge (1.022 MeV), $^{85}$Sr (0.514 MeV), $^{137}$Cs (0.661 MeV), and 
	$^{203}$Hg (0.279 MeV). The effects of scintillation quenching, Cherenkov light production, and PMT dark hits on the
	energy scale non-linearity are determined from this calibration data.
	They are corrected for in the spectral analysis discussed later.
	The observed vertex resolution is $\sim$13\,${\rm cm}/\sqrt{E{\rm (MeV)}}$, and 
	the energy resolution ($\sigma_E/E$) is $(6.9 \pm 0.1)\%/\sqrt{E{\rm (MeV)}}$.  
	The deviation of the position dependent energy from the center of the detector is 
	evaluated as $+0.3$\% to $-1.5\%$ inside of the $-4.5 < z < 4.5{\rm~m}$ region shown in Figure \ref{figure:z-axis-calibration}.
        The majority of the calibrations were performed by moving the sources to specific points along
	the central vertical detector axis using the deployment system described in Ref.~\cite{Banks2015}.  In order to constrain deviations from
	rotational symmetry after scintillator purification, a full 3D-calibration was 
	performed using an off-axis system described in Ref.~\cite{Berger2009}.  	

%

\section{Event Selection}
\label{section:EventSelection}

	\vspace{-0.2cm}

	Candidate events are selected according to the following requirements:
	
	(i) The radial position of event vertices must be less than 4.5 m.
	An additional volume selection, defined in cylindrical coordinates and motivated by the data classification described in the next section, is also applied. 
	The combined volume selection defines the fiducial volume (FV) which corresponds to 268.6 tons of liquid scintillator.

	(ii) Cosmic-ray muons (identified by a total PMT charge of 
	larger than 10,000 photo electrons (p.e.) or more than 5 PMT hits in the outer detector) and all 
	events within 2\,ms after muons are rejected to reduce background events 
	due to muon spallation products and electronics noise.
	In addition, noise events within 100\,$\mu$s after high energy events 
	(a total PMT charge of larger than 1,000 p.e.) are rejected.
	
	(iii) Two successive events within 1\,$\mu$s are rejected 
	to avoid the possible cross talk effect between two events 
	owing to the finite time spread of scintillation photons.
	
	(iv) Coincidence events occurring within 1.2 ms of each other are eliminated
	in order to remove $^{214}$Bi-$^{214}$Po and $^{212}$Bi-$^{212}$Po sequential decays.
		
	(v) Candidates must pass a vertex-time-charge (VTQ) fit quality test to eliminate noise 
	events mainly produced by two-event pile-up in a one-event time window ($\sim$200 ns). The 
	VTQ cut is tuned using calibration data. The reduction of the neutrino event selection 
	efficiency is found to be negligible in the analysis energy range.

%

\section{Data Classification}
\label{section:DataRankSection}

	\vspace{-0.2cm}

\begin{figure}[t]
		\begin{center}
			\includegraphics[angle=0,width=1.0\columnwidth]{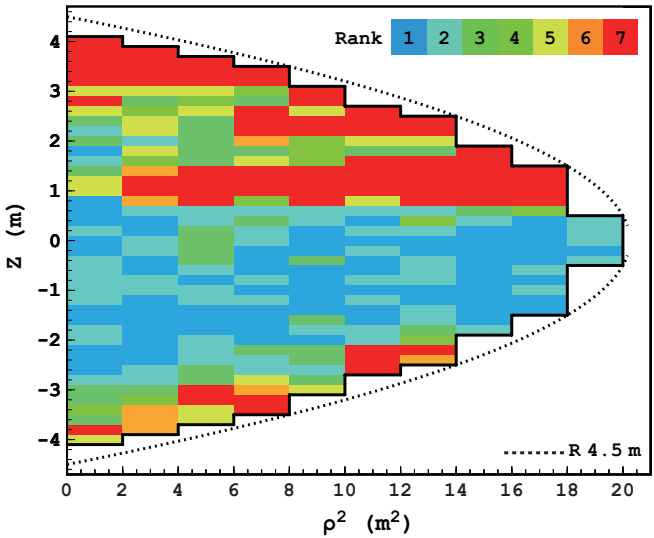}
			\vspace{-0.8cm}
		\end{center}
		\caption[]{"Color" Rank classification of partitions in the fiducial volume during a short data 
		taking interval.}
		\label{figure:classification}
	\end{figure}

	After the introduction of purified LS into KamLAND there were time periods of thermal 
	instability due to slight variations in the temperature gradient of the detector.  
	It was further found that the containment balloon acts as a $^{210}$Pb reservoir,
	slowly releasing $^{210}$Bi into the scintillator.
	The result of these thermal gradients was convection in the LS fiducial volume and a non-uniformly distributed $^{210}$Bi 
	concentration, thus some regions of the fiducial volume were much cleaner than others.  
	Choosing to only analyze regions of the fiducial volume that contain low concentrations 
	of $^{210}$Bi could introduce a selection bias. Thus, a procedure for analyzing all the 
	data regardless of the local $^{210}$Bi concentration was developed.  The procedure used in this
	analysis is as follows:
	\begin{enumerate}
		\item From the perspective of a $\rho^{2}$ vs.~$z$ distribution in cylindrical coordinates, 
		where $\rho^{2} = x^{2}+y^{2}$,
		the fiducial volume is divided into equal-volume partitions having dimensions 
		$d\rho^{2}= 2.0$\,m$^{2}$ and $dz = 0.2$\,m.
		\item For events with an energy of 0.5$-$0.8\,MeV (mainly $^{210}$Bi) 
		an effective event rate is calculated for each partition. Each partition is assigned
		a rank based on the average rate of its spatial and temporal neighbors. 
		An average rate is calculated for up to eight partitions bordering the cell to be ranked 
		(a partition near the FV boundary has fewer neighbors). 
		The time dimension is included in this by adding the rate of the ranked cell, 
		as determined in the previous and following data taking runs (usually one day long), to the average rate of its neighbors.
		\item The effective event rate is then used to 
		classify each partition into one of seven different {\it ranks} which are defined in Table~\ref{table:ranks}.  
		As an example, the rank classification for one data-taking run ($\sim$ 24 hours long) is shown in Figure\,\ref{figure:classification}.  
		We find the rank assignment of each volume generally varies slowly with time except after periods of thermal instability 
		when convection occurred in the LS.
	\end{enumerate}

	\begin{center}
		\begin{table}[b!]
			\caption{\label{table:ranks}Listing of the ranks defined according to the estimated $^{210}$Bi rate in 
			$(0.5<E<0.8\,{\rm MeV})$ and the total exposure for each rank.
			}
			\begin{tabular}{@{}*{2}{lcc}}
				\hline\hline
				Rank & ~~~~$^{210}$Bi Rate~~~~ & Exposure \\
				\# & $10^{-6}$(m$^{3}$ s)$^{-1}$ & (kton-days)\\
				\hline
				1 &  $< 5$ & 26.52\\
				2  &  $5-10$ & 34.42\\
				3  &  $10-15$ & 27.06\\
				4 &  $15-20$ & 17.81\\
				5 &  $20-25$ & 11.35\\
				6 & $25-30$ & 7.63\\
				7 & $> 30$ & 40.63\\
				\hline
				Total & &  165.43\\
				\hline\hline
			\end{tabular}
		\end{table}
	\end{center}

	The procedure just outlined allows identification of low $^{210}$Bi regions within the detector without
	complex and arbitrary fiducial volume cuts. 
        However, each partition will have some volume bias due to vertex resolution and position-dependent
	vertex reconstruction.  This bias is corrected in the spectral fit using the fact that the following
	event rates, produced from cosmogenic or astrophysical sources, must be distributed uniformly in the LS:  
	$^{11}$C, $^{10}$C, $^{7}$Be and $^{7}$Be solar neutrinos.
        Using the prescribed rank classification, a simultaneous spectral fit is performed over 
	the data from every rank to obtain one common $^{7}$Be solar neutrino rate.

%

\section{Background Estimation}
\label{section:Background}

	\vspace{-0.2cm}

	Accurate modeling of the energy distributions of background sources inside and outside the detector is necessary to determine the
	$^{7}$Be solar neutrino flux with KamLAND. These energy distributions are based
	on a phenomenological detector response function. They derive their validity from the fact
	that they describe the data well (a) before purification when radio-impurity 
	concentrations were high, and (b) in areas of high rate after purification.
	For some sub-dominant components, that cannot be verified in this way, Monte
	Carlo generated spectra are used instead. The contributions of the various background components are summed 
	with freely-varying normalizations, although some normalizations are constrained by other independent KamLAND data.  
	A fit to the candidate event spectrum then determines the partial contributions.  
	
	This section describes how the background model
	is constructed and what is known about its constituents.
	Background sources are classified into three categories: radioactive impurities 
	in the LS, spallation products, and radioactivity in the surrounding materials. 
	\begin{figure}[t]
		\begin{center}
			\includegraphics[angle=0,width=1.0\columnwidth]{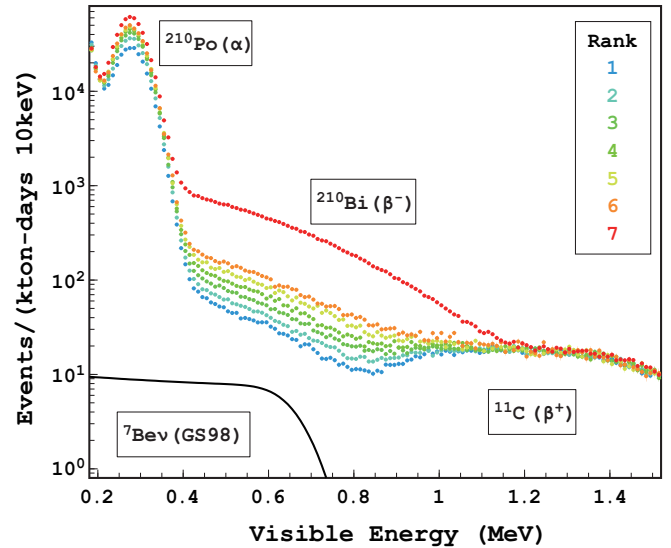}
			\vspace{-0.8cm}
		\end{center}
		\caption[]{"Color" Energy spectra of candidate events for each data-set rank, exhibiting different $^{210}$Bi 
		rates. All data sets are used for the simultaneous fit of the $^{7}$Be solar 
		neutrino spectrum, with a common rate normalization.
		The location of the major background sources are labeled and the theoretical electron recoil spectrum of 862 keV $^7$Be
		solar neutrinos, whose rate is based on the standard solar model by Serenelli et al.~\cite{Serenelli2011}, is shown
		for comparison.}
		\label{figure:energy_rank}
	\end{figure}

	\subsection{Radioactive Impurities in the Liquid Scintillator}
	\label{section:Background_RadioactiveImpurities}
		
		The abundance of $^{232}$Th and $^{238}$U and their daughters in the LS is constrained by KamLAND data. 
		The decay rates of the $^{222}$Rn-$^{210}$Pb subchain of the $^{238}$U series 
		and the $^{228}$Th-$^{208}$Pb subchain of the $^{232}$Th series can be measured, almost background free, 
		using $^{214}$Bi-$^{214}$Po and $^{212}$Bi-$^{212}$Po beta-alpha delayed coincidences, 
		with 94.8\% and 31.9\% selection efficiencies, respectively. 
		The inferred $^{238}$U and $^{232}$Th concentrations assuming secular equilibrium are $(5.0 \pm 0.2) \times 10^{-18}$\,g/g
		($93 \pm 4$\,nBq/m$^3$) and $(1.3 \pm 0.1) \times 10^{-17}$\,g/g ($59 \pm 4$\,nBq/m$^3$), respectively.
		Some radioactive decay chains are found to be out of
		secular equilibrium. A fit to the prominent $^{210}$Po $\alpha$-peak,
		quenched to about 0.3 MeV visible energy in Figure~\ref{figure:energy_rank},
		yields spatially varying activities ranging from 2.4\,mBq/m$^3$ to
		4.8\,mBq/m$^3$, depending on the rank of the analyzed volume.
		However, as discussed in the following subsection, this background component is not considered 
		in the fit since it falls below the analysis threshold. 
		The higher-mass members of the decay chains
		above $^{228}$Th and $^{222}$Rn are determined by a spectral fit and 	are found to be unimportant.
		
		Below 1 MeV, the background is dominated by daughters of $^{210}$Pb, namely $^{210}$Bi 
		and $^{210}$Po. $^{210}$Bi, the most important background contributor,
		is found to be spatially non-uniform, in addition, its decay rate in the FV fluctuates in time.
		This is attributed to $^{210}$Bi supplied from the balloon surface by irregular 
		LS convection currents, followed by its decay with $T_{1/2} = 5.01\,{\rm days}$. 
		This interpretation is supported by the observation that around the balloon surface, 
		the spectrum is dominated by electrons from $^{210}$Bi decays, and $\alpha$-particles from 
		$^{210}$Po decays, consistent with the hypothesis that the balloon film 
		is contaminated with $\sim$200\,Bq of $^{210}$Pb, a $^{222}$Rn daughter introduced during detector construction.
		The scintillator sub-volume ranking technique, described before, was devised
		to cope with this variability in an unbiased way.  Depending on the rank, the
		specific $^{210}$Bi activity varies between 35.7\,$\mu$Bq/m$^3$ and 681.1\,$\mu$Bq/m$^3$
		within the FV. $^{210}$Bi undergoes a first forbidden beta decay. The parametrization
		of the resulting beta spectrum will be discussed in the analysis section.

		The concentration of $^{40}$K is measured to be $(7.3 \pm 1.2) \times 10^{-17}$\,g/g from the energy 
		spectrum fit. Due to the fluid circulation during LS purification, $^{238}$U, $^{232}$Th, 
		and $^{40}$K may have non-uniform spatial distributions.
		The long-lived radioactive noble gas, $^{85}$Kr, was the major source of low-energy
		background prior to scintillator purification. The present background model
		includes this component to deal with any left-over activity. $^{39}$Ar is included
		as a potential source in the background model, its abundance constrained by the ratio of $^{85}$Kr and
		$^{39}$Ar found in the atmosphere. From the spectral fit we find its contribution to the background is negligible.
		This is consistent with the pre-purification spectral analysis results.
		For volume ranks 1 through 6 the $^{85}$Kr specific activity varies
		between 4.1\,$\mu$Bq/m$^3$ and 19.2\,$\mu$Bq/m$^3$. For the highest background partitions --- those with 
		rank 7 --- the fit only yields an upper limit for the activity.
		After LS purification, we collected samples from a location within the fiducial 
		volume at z = +1.5\,m. These samples were analyzed for their Kr content using a helium
		purge, a cold trap to retain the Kr, and a residual gas analyzer to measure its	partial pressure.
		Based on these measurements, the $^{85}$Kr decay rate was estimated at $8.3 \pm 4.2\,\mu{\rm Bq} 
		/ {\rm m}^{3}$, assuming a recent isotopic ratio of $^{85}$Kr in air~\cite{Hirota2004}. 
		As a third cross check for the $^{85}$Kr content of the LS, a $^{85}$Kr-$^{\rm 85m}$Rb
		delayed coincidence analysis was performed, utilizing the beta decay into the
		metastable, 514 keV excited state of $^{85}$Rb. This analysis yielded 
		$17.3 \pm 5.9\,\mu {\rm Bq} / {\rm m}^{3}$, averaged over the entire FV.
		$^{85}$Kr undergoes a unique first forbidden beta decay. The beta spectrum
		contained in the background model was calculated using a relativistic
		Fermi function plus a shape correction accounting for the forbiddenness following
		the procedure outlined in Ref.~\cite{Konopinski1966}.
		This spectral parametrization was found to fit well the high statistics beta spectrum
		collected before LS purification.

	\subsection{Spallation Products}
	\label{section:Background_SpallationProducts}
	
		The 2700 m.w.e.~of rock overburden of the Kamioka Underground Laboratory
		suppresses the rate of cosmic ray muons traversing the KamLAND LS to \mbox{$0.198\pm0.014$\,s$^{-1}$}
		~\cite{Abe2010}. The surviving muons can produce unstable light nuclei by spallation of carbon, 
		whose decays result in background. 
		The dominant cosmogenic background between 1-2\,MeV 
		is due to decays of $^{11}$C ($\beta^{+}$, \mbox{$\tau=29.4$\,min}, 
		\mbox{$Q=1.98$\,MeV}). Due to its relatively long half life these decays cannot be
		tagged without incurring large deadtime.
		Its decay rate has earlier been estimated to be \mbox{$1106\pm178\,($kton-days$)^{-1}$}
		~\cite{Abe2010}. The background contribution of $^{11}$C is constrained by this value.
		
		Another spallation source of interest is 
		$^{7}$Be (EC decay, \mbox{$\tau=76.8$\,days}, \mbox{$Q = 0.862$~MeV}). 
		While its production rate by muon spallation in the LS is estimated to be small~\cite{Hagner2000, Abe2010}, 
		there is the possibility of higher than steady state production yield due to the introduction of fresh, surface-exposed 
		LS during the scintillator purification. Therefore, in order to be conservative, the rate of $^{7}$Be decays is unconstrained.

	\subsection{Radioactivity in the Surrounding Material}
	\label{section:Background_ExternalRadioactivities}

		The background from external gamma-rays is mainly caused by $^{40}$K, $^{232}$Th, and $^{238}$U contained 
		in the surrounding rock, stainless steel, PMT glass, balloon film, and Kevlar suspension ropes. 
		The energy distributions resulting from these radiation sources were modeled by means of a Monte Carlo simulation.
		The simulation was tuned with source calibration data to reproduce the vertex distribution as well.
		The gamma-ray attenuation in the radial direction in the simulation is consistent with that in the real data.
		External backgrounds dominate the energy distribution for radial positions larger than 4.5\,m. These data
		were used to fit the relative background contributions. The Monte Carlo simulation was then used to
		extrapolate the background for $R < 4.5\,{\rm m}$.
		Based on this fit, it was concluded that external gamma-rays do not significantly impact the fiducial volume background below 1\,MeV.
		
		As discussed before, the balloon surface is a source of electrons from $^{210}$Bi decays, and $\alpha$-particles 
		from $^{210}$Po decays. While the fiducial volume cut effectively suppresses these backgrounds, accidental pile-up of two 
		external events can lead to a vertex and energy displacement, moving external events into the analysis volume.
		Due to the high $^{210}$Po decay rate, resulting in a pronounced peak at $\sim$0.3\,MeV in the energy spectrum 
		(see Figure~\ref{figure:energy_rank}), most of the pile-up events are concentrated in the high energy tail
		of this peak.  Although the absolute rate of tail events is not small, the fraction of pile-up events to observed events 
		above 0.5\,MeV is $<$1\%.
		To eliminate this background and the systematic bias it would introduce into the measurement 
		of the $^{7}$Be solar neutrinos, the energy threshold was set to 0.5\,MeV in the present analysis.
		\begin{center}
			\begin{table}[b]
				\caption{\label{table:systematic}Uncertainties on the measurement of the $^{7}$Be 
				solar neutrino flux.}
				\begin{tabular}{@{}*{2}{lc}}
					\hline\hline
					Source & Uncertainty (\%) \\
					\hline
					Cross section & 1.0 \\
					Number of target & 0.10 \\
					Fiducial volume & 3.4 \\
					Vertex misreconstruction &  0.5 \\
					Energy scale & 7.9 \\
					Rank-dependent energy scale & 2.9 \\
					Energy resolution & 3.4 \\
					BG from $^{238}$U-series ($^{222}$Rn-$^{210}$Pb) & 1.7 \\
					BG from $^{232}$Th-series ($^{228}$Th-$^{208}$Pb) & 1.8 \\
					BG from other solar neutrinos & 1.9 \\
					\hline
					Systematic total & 10.2 \\
					\hline
					Statistics & 12.4 \\
					\hline
					Total & 16.1\\
					\hline\hline
				\end{tabular}
			\end{table}
		\end{center}

%

\vspace{-0.55cm}
\section{Systematic Uncertainty}
\label{section:SystematicUncertainty}
	The leading contributions to the systematic uncertainty of the $^{7}$Be solar neutrino flux measurement 
	are listed in Table~\ref{table:systematic}. 
	The measured neutrino-electron scattering rate is converted into a solar neutrino flux, 
	the accuracy of this conversion is given by the uncertainty of the interaction cross section.
	Based on the evaluation of Ref.~\cite{Bahcall1995}, a value of 1\% is assigned to this error. 
	The determination of the flux further requires knowledge of the number of electrons contained 
	in the FV. The LS density is measured to be 0.780\,${\rm g}/{\rm cm}^{3}$ with an 
	uncertainty of 0.025\% at 11.5$^{\circ}$C. The uncertainty of the dependence of the number of 
	electrons on the temperature within the FV is estimated at 0.1\%. 
	We estimate that $9.21 \times 10^{31}$ electrons are contained in the 344.3\,${\rm m}^{3}$ FV.
	Data collected with the full volume calibration system showed vertex reconstruction 
	deviations of less than 5\,cm. This corresponds to a FV uncertainty of 3.4\%.
	The FV event selection inefficiency due to vertex misreconstruction was established with source calibrations and is less than 0.5\%
	in the analysis energy region, confirmed by the source calibrations.
	
	Other systematic uncertainties, related to the modeling of the detector response, are 
	determined through the spectral fit procedure, which is presented in the following section.
	The correction for the non-linearity of the energy scale for each particle type
	($\gamma$, $e^{-}$ and $e^{+}$) is performed by varying the energy response parameters in the $\chi^{2}$-fit. 
	The fit model, on which the solar neutrino analysis is based, uses free-floating energy scales with constraints 
	from calibration data
	in the form of penalty terms.
	As such, the fit uncertainty already includes the energy scale uncertainty. 
	In order to quantify the contribution of the energy scale uncertainty as a separate item in the error budget,
	we repeat the fit with the energy scale parameters fixed. 
	The energy scale error is stated as the quadratic difference between the errors obtained from free-floating and fixed parameter fits.
	This method implies a 7.9\% uncertainty on the best-fit $^{7}$Be rate due to the uncertainty of  the energy scale.
	To check the possibility of the energy scale varying with the rank we compared the fit result for two extreme cases, namely
	fully correlated and fully independent energy scale between ranks.  We adopt the deviation of the best-fit $^{7}$Be
	as an estimate of the associated systematic uncertainty.  To propagate the $0.1\%/\sqrt{E{\rm (MeV)}}$ uncertainty
	of the energy resolution we repeat the analysis varying the detector resolution within this uncertainty.  We find that this
	causes the best-fit $^{7}$Be rate to vary by 3.4\%.
	
	The definition of rank boundaries (Table \ref{table:ranks}) and their effect on the best-fit $^{7}$Be rate
	was also studied.  It is important to note that the fit sensitivity to the solar signal comes from ranks 1 and 2 where the 
	signal-to-background is largest, hence changing the boundary definitions of the other ranks has negligible effect. 
	The fitting procedure was repeated while varying the boundary between ranks 1 and 2 from $3\times10^{-6}$ to $7\times10^{-6}$ 
	events/(m$^{3}$ s).  
	The choice of $5\times10^{-6}$ events/(m$^{3}$ s) as the boundary was found to have the highest fit probability,
	quantified by $\chi^{2}/$n.d.f, and is used in the final analysis presented here.
	Furthermore, we find the best-fit rate for all rank boundaries considered  were consistent with each other to within $5\%$ 
	and conclude that the choice of boundary does not bias the result.

	The largest background contribution, $^{210}$Bi first forbidden beta decay, has an 
	additional uncertainty related to its shape correction. Using our data, mainly rank 7 where the background rate 
	is highest, we derived a phenomenological fourth-order polynomial correction to the shape factor published in~\cite{Daniel1962}
	to better model the $^{210}$Bi shape.  
	The parameters of this correction polynomial ($a_{n}$) are: 
	$a_0=1$, $a_1 = 41.4 \pm 0.8$, $a_2 = -101.2 \pm 1.4$, $a_3 = 102.9 \pm 3.5$, and $a_4 = -37.9 \pm 2.0$ in units of MeV$^{-n}$. 
	Within the analysis
	energy window, deviations of the spectral shape from that of Ref.~\cite{Daniel1962} were
	small.
	
	Position dependent biases in vertex reconstruction could introduce an artificial non-uniformity
	in the distribution of $^{7}$Be solar 
	neutrinos within the FV, resulting in a systematic error on the simultaneous fit to different 
	rank data. The effects of the non-uniformity are parameterized, 
	and constrained for each rank 
	spectrum to obtain uniformly distributed $^{11}$C events.
	
	The systematic uncertainties due to the $^{238}$U and $^{232}$Th series and other solar neutrinos
	are evaluated by the deviation of the $^7$Be solar neutrino rate on varying their rates within $\pm 1\sigma$.
	\begin{center}
		\begin{table}[t]
			\caption{\label{table:background}Summary of signal and background in the fiducial volume. 
			The best-fit signal and background rates in the whole energy range and the $^7$Be solar neutrino energy range 
			($0.5\,{\rm MeV} < E  <0.8\,{\rm MeV}$) are shown.
			}
			\begin{tabular}{@{}*{3}{ccccc}}
				\hline\hline
				Isotope & \multicolumn{2}{c}{Event rate  w/o $E$-cut} &  \multicolumn{2}{c}{Event rate in 0.5-0.8\,MeV} \\
				 & all ranks & rank-1 & all ranks & rank-1 \\
				& \multicolumn{2}{c}{$($kton-days$)^{-1}$ } &  \multicolumn{2}{c}{$($kton-days$)^{-1}$ } \\
				\hline
				\\
				\multicolumn{5}{c}{Solar neutrinos} \\
				\hline
				$^{7}$Be $\nu$ 	&  \multicolumn{2}{c}{$582 \pm 94$} 	& \multicolumn{2}{c}{$117 \pm 19$}\\
				Other $\nu$		&  \multicolumn{2}{c}{$1443$}			& \multicolumn{2}{c}{$14.9$} \\
				\hline
				\\
				\multicolumn{5}{c}{Radioactive impurities in the LS} \\
				\hline
				$^{210}$Bi	&  $23974 \pm 883$ 	& $3955 \pm 238$ 	& $4557 \pm 168$	& $578 \pm 35$\\
				$^{85}$Kr 	&  $858 \pm 59$ 		& $453 \pm 102$ 	& $57 \pm 4$ 		& $30 \pm 7$\\
				$^{39}$Ar 	&  $3 \pm 3$ 			& $2 \pm 3$ 		& $0.04 \pm 0.04$ 	& $0.02 \pm 0.04$\\
				$^{40}$K 		&  $181 \pm 29$ 		& $3 \pm 38 $ 		& $48 \pm 8$		& $1 \pm 10$\\
				\hline
				\\
				\multicolumn{5}{c}{Spallation products} \\
				\hline
				$^{7}$Be	&  \multicolumn{2}{c}{$167 \pm 173$} 	& \multicolumn{2}{c}{$2 \pm 2$}\\ 
				$^{11}$C	&  \multicolumn{2}{c}{$973 \pm 10$} 	& \multicolumn{2}{c}{$3.55 \pm 0.04$}\\ 
				Other 	&  \multicolumn{2}{c}{$173$}			& \multicolumn{2}{c}{$8.0$}\\
				\hline
				\\
				\multicolumn{5}{c}{Others} \\
				\hline
				external-$\gamma$ 	&  --- & --- & $2.4 \pm 0.4$ 	& $2 \pm 1$\\
				pile-up 			&  --- & --- & $10 \pm 1$ 	& $8 \pm 2$\\
				\hline\hline
			\end{tabular}
		\end{table}
	\end{center}

%

\section{Result}
\label{section:Result}

	\vspace{-0.2cm}

	The $^{7}$Be solar neutrino rate is estimated by a likelihood fit to the binned energy spectra of 
	candidate events with visible energy between 0.5 and 1.4\,MeV. The background contributions from $^{210}$Bi, $^{85}$Kr, 
	$^{40}$K, and $^{7}$Be EC are free parameters and their normalizations are left unconstrained in the fit. 
	The contributions from the $^{222}$Rn-$^{210}$Pb and $^{228}$Th-$^{208}$Pb chains, and $^{11}$C are 
	allowed to vary but their normalizations are constrained by independent KamLAND measurements,
	as outlined in Section~\ref{section:Background}.
	Distributions with different background rank are fitted simultaneously. 
	The background rates, derived from the fit normalizations, are summarized in Table~\ref{table:background}. 
	The backgrounds from external gamma-rays, 
	and pile-up events are constrained by the MC study. 
	The model parameters of the detector response are also constrained,
	as discussed in Section~\ref{section:SystematicUncertainty}.
	
\begin{figure}[t!]
		\begin{center}
			\includegraphics[bb=0 0 830 573, angle=0,width=1\columnwidth]{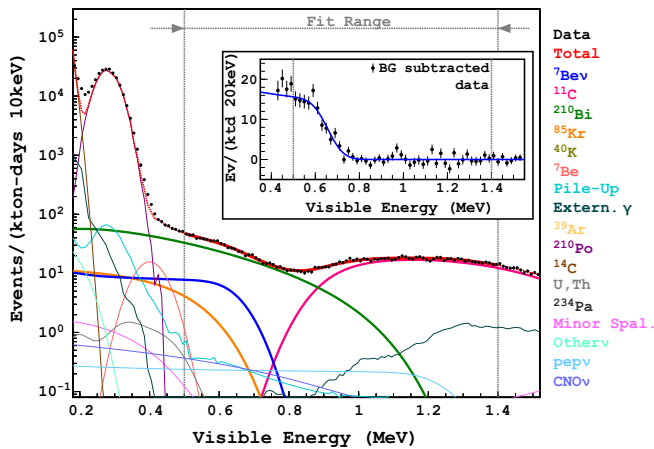}
			\vspace{-0.8cm}
		\end{center}
		\caption[]{"Color" Energy spectrum of selected $^{7}$Be solar neutrino candidates for the rank-1 data-set together with the 
		best-fit curves from the simultaneous fit to all rank's spectra. The fit range is $0.5\,{\rm MeV} < E < 1.4\,{\rm MeV}$. 
		The background-subtracted spectrum is shown in the inset, 
		where a Compton-like shoulder characteristic for the $^{7}$Be solar neutrino contribution is evident.
		The error bars are statistical only and do not include correlated systematic uncertainties in the 
		background rate estimates.}
		\label{figure:energy_fit}
	\end{figure}

\begin{figure}[h!]
		\begin{center}
			\includegraphics[bb=0 0 655 316, angle=0,width=0.9\columnwidth]{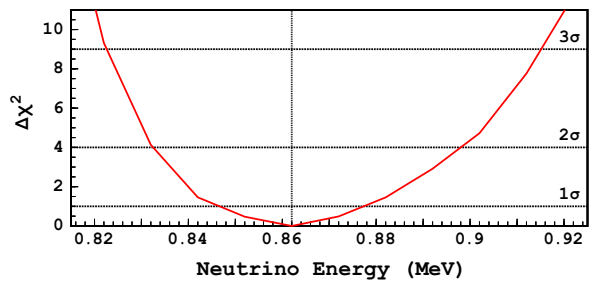}
			\vspace{-0.8cm}
		\end{center}
		\caption[]{"Color Online" $\Delta \chi^{2}$-profile from the fit to the energy of mono-energetic neutrinos discussed in the text. 
		The vertical line indicates the neutrino energy predicted from the $^{7}$Be electron-capture reaction at 862\,keV.}
		\label{figure:chi2}
	\end{figure}

	Figure~\ref{figure:energy_fit} shows the result of this procedure: the best-fit spectrum for the rank-1 
	data-set which is lowest in background, and therefore, most sensitive to the solar neutrino
	signal. The inset shows the background subtracted energy distribution, which 
	exhibits the shape of the best-fit $^7$Be neutrino signal extracted from the analysis.
	The backscattering edge is visible and located at the correct energy.
	The $\chi^{2}$/d.o.f. comparing the binned data and the 
	best-fit model is found to be 635.3/589 using data of all ranks.
	The $\chi^{2}$/d.o.f. of the lowest background rank-1 data is 80.1/90.
	As can be seen from the data in Table~\ref{table:background} the solar-signal to background ratio 
	is estimated to be
	1:5.5 for rank 1 data ($0.5<E<0.8\,{\rm MeV}$). 
	This relatively low signal to background ratio naturally invites the question whether
	a statistically significant $^7$Be solar neutrino signal is really present in the data.
	The fit to the data was also performed assuming the absence of a solar neutrino signal,
	but minimizing all other background components within their constraints as before. The observed
	increase in the fit $\chi^2$ leads us to reject the no-solar signal hypothesis
	at 8.2$\sigma$ CL. 
	To understand whether the solar recoil signal, preferred by the data and shown in the inset of Figure\,\ref{figure:energy_fit}, 
	is a unique solution or whether  other continuous distributions yield equally good fits, 
	the neutrino energy (assuming a mono-energetic source) and interaction rates were floated 
	and the fit $\chi^2$-profile determined under their variation. 
	The result of this procedure is shown in Figure\,\ref{figure:chi2}.
	The KamLAND data clearly prefer the presence of an edge at an energy
	that coincides with that expected for solar $^7$Be neutrino induced electron recoils.
	Assuming that the edge is indeed solar neutrino induced, we determine the solar
	neutrino energy to be $E_{\nu} = 862 \pm 16\,{\rm keV}$. This is the first direct spectroscopic
	determination of the solar $^7$Be neutrino energy.

	The fit to the KamLAND data gives a solar neutrino interaction rate of
	\mbox{${\rm R}_{\rm KL}=582 \pm 94\,($kton-days$)^{-1}$}. The quoted error corresponds to the quadratic sum of the
	statistical and systematic errors, listed in Table~\ref{table:systematic}.
	This result is in agreement with the latest interaction rate reported by the
	Borexino experiment: \mbox{${\rm R}_{\rm B}=460^{+21}_{-22}\,($kton-days$)^{-1}$}~\cite{Bellini2013}.
	The rate difference (${\rm R}_{\rm KL}-{\rm R}_{\rm B}$) deviates by $1.3\sigma$ from zero.
	Differences in the chemical composition of the liquid scintillators used in both experiments result in a 3.6\% difference of the rates.
	Assuming that the ES interactions detected by KamLAND are due to a pure electron flavor 
	flux, KamLAND's interaction rate corresponds to a 862\,keV $^{7}$Be solar neutrino flux of 
	$(3.26 \pm 0.52) \times 10^{9}\,{\rm cm}^{-2}{\rm s}^{-1}$.
	The standard solar model (SSM) by Serenelli et al.~\cite{Serenelli2011}
	gives two $^7$Be solar neutrino flux values, depending whether the older Grevesse and Sauval
	(GS98)~\cite{Grevesse1998} or the Asplund (AGSS09)~\cite{Asplund2009} solar abundances are utilized.
	When the GS98 solar abundances are assumed, the SSM flux value is 
	$(5.00 \pm 0.35) \times 10^{9}\,{\rm cm}^{-2}{\rm s}^{-1}$. The flux reduces to
	$(4.56 \pm 0.32) \times 10^{9}\,{\rm cm}^{-2}{\rm s}^{-1}$ under the assumption of the AGSS09 solar abundances.
	Assuming that the $\nu_{e}$ mix with $\nu_{\mu}$ and $\nu_{\tau}$ the KamLAND flux
	measurement corresponds to a survival probability, $P_{ee} = 0.66 \pm 0.15$.
	This value was obtained taking into account the cross section difference for these 
	neutrino flavors. 
	
	Using KamLAND's 2013 best-fit neutrino oscillation parameters~\cite{Gando2013}, 
	based on a global oscillation analysis under the assumption of $CPT$-invariance, the 
	survival probability is better constrained. In this case, we obtain a total $^{7}$Be neutrino flux of 
	$(5.82 \pm 1.02) \times 10^{9}\,{\rm cm}^{-2}{\rm s}^{-1}$. 
	This result is consistent with the flux determination provided by Borexino: 
	$(4.75 ^{+0.26}_{-0.22}) \times 10^{9}\,{\rm cm}^{-2}{\rm s}^{-1}$~\cite{Bellini2013},
	and while it somewhat favors the GS98 model flux, the KamLAND data are not precise enough to
	discriminate between the two solar model fluxes.

%

\section{Conclusion}
\label{section:Conclusion}

	\vspace{-0.2cm}

	The KamLAND collaboration reports a new measurement of the $^{7}$Be solar neutrino interaction rate in
	a liquid scintillator. Performing this difficult measurement required an extensive purification
	campaign, reducing the scintillator's radio-impurity content by several orders of magnitude
	The measured $^7$Be solar neutrino rate in KamLAND is \mbox{${\rm R}_{\rm KL}=582 \pm 94\,($kton-days$)^{-1}$},
	which corresponds to  a flux of $(5.82 \pm 1.02) \times 10^{9}\,{\rm cm}^{-2}{\rm s}^{-1}$ under the assumption of 
	KamLAND's 2013 best-fit oscillation parameters.

	The statistical significance of this signal is estimated to be 8.2$\sigma$ and provides the first
	independent verification of the only prior measurement of this quantity, performed by Borexino.
	The solar neutrino flux derived from the KamLAND data agrees with the solar model
	values, but is not accurate enough to shed light on the question of solar metallicity.

~
%

\section*{ACKNOWLEDGMENTS}

	\vspace{-0.2cm}

	The \mbox{KamLAND} experiment is supported by the Grant-in-Aid for Specially Promoted Research under 
	grant 16002002 and 21000001 of the Japanese Ministry of Education, Culture, Sports, Science and Technology; 
	the World Premier International Research Center Initiative (WPI Initiative), MEXT, Japan; 
	Stichting FOM in the Netherlands; and under the US Department of Energy (DOE) 
	Grant No. DE-AC02-05CH11231 and DE-FG02-01ER41166, as well as other DOE grants to individual institutions. 
	The Kamioka Mining and Smelting Company has provided service for activities in the mine.

\bibliography{SolarNeutrino}
\end{document}